\documentclass[reprint,twocolumn,showpacs,superscriptaddress,groupedaddress]{revtex4}  % for review and submission
\usepackage{graphicx}  % needed for figuresp
\usepackage{dcolumn}   % needed for some tables
\usepackage{bm}        % for math
\usepackage{amssymb}   % for math
\usepackage{color}   % for color
\usepackage{amsmath}
\usepackage{amsfonts}
\usepackage{upgreek}
\usepackage{ulem}
\usepackage{caption}
\usepackage{subfig}

\newcommand{\beginsupplement}{%
        \setcounter{table}{0}
        \renewcommand{\thetable}{S\arabic{table}}%
        \setcounter{figure}{0}
        \renewcommand{\thefigure}{S\arabic{figure}}%
        \setcounter{equation}{0}
        \renewcommand{\theequation}{S\arabic{equation}}%
     }

\begin{document}

\title{Effect of the polydispersity of a colloidal drop on the drying induced stress as measured by the buckling of a floating sheet}

\author{Fran\c{c}ois Boulogne}
%\email[]{Your e-mail address}
\affiliation{Department of Mechanical and Aerospace Engineering, Princeton University, Princeton, NJ 08544}
\author{Yong Lin Kong}
%\email[]{Your e-mail address}
\affiliation{Department of Mechanical and Aerospace Engineering, Princeton University, Princeton, NJ 08544}
\author{Janine K. Nunes}
%\email[]{Your e-mail address}
\affiliation{Department of Mechanical and Aerospace Engineering, Princeton University, Princeton, NJ 08544}
\author{Howard A. Stone}
%\email[]{Your e-mail address}
\affiliation{Department of Mechanical and Aerospace Engineering, Princeton University, Princeton, NJ 08544}

\date{\today}

\begin{abstract}
    We study the stress developed during the drying of a colloidal drop of silica nanoparticles.
    In particular, we use the wrinkling instability of a thin floating sheet to measure the net stress applied by the deposit on the substrate and we focus on the effect of the particle polydispersity.
    In the case of a bidisperse suspension, we show that a small number of large particles substantially decreases the expected stress, which we interpret as the formation of lower hydrodynamic resistance paths in the porous material.
    As colloidal suspensions are usually polydisperse, we show for different average particle sizes that the stress is effectively dominated by the larger particles of the distribution and not by the average particle size.
\end{abstract}

\maketitle

Drying of colloidal droplets is ubiquitous in processes such as ink-jet printing technologies \cite{James2011} or spray painting.
Recently, new directions for printing techniques have been developed for soft materials to enable applications such as conformable electronics, soft robotics and wearable devices.
For instance, electronic circuits can be printed on elastomers \cite{Rogers2010,Boley2015} and human skin \cite{Jeong2013}.
However, deformations of the surface can be induced by the surface tension of a liquid drop \cite{Jerison2011,Marchand2012,Dervaux2015} or the consolidation of a colloidal material \cite{Boulogne2014b}, which could affect the final quality and the function of the printed device.
Here, we report measurements of the drying induced stress and highlight the role of polydispersity of the suspension.

Previous reported measurements of drying or cracking stresses rely on the deflection of a cantilever beam \cite{Petersen1999,Yow2010,Thomas2011,Chekchaki2013} where a liquid film of a polymer solution or colloidal suspension is coated on a thin flexible plate.
In such a geometry, the film dries from the edge toward the center of the film \cite{Li2012a}, which leads to spatially inhomogeneous states of the material.

However, besides the bending, when a stress is applied above a certain limit to a thin sheet, a wrinkling instability can be observed \cite{Cerda2003,Geminard2004,Oshri2015,Damman2015}.
Recent studies focused on an elastomeric disk floating on a liquid bath with a liquid drop in the center of the disk \cite{Huang2007,Davidovitch2011,King2012,Toga2013,Pineirua2013}.
In these situations, the stress applied to the membrane is due to the surface tension of the involved liquids.
Therefore, these studies of a deformable sheet established  models based on F\"oppl-von K\`arm\`an equations to predict the length and the number of wrinkles.

\begin{figure}[h!]
    \centering
    \includegraphics[width=.9\linewidth]{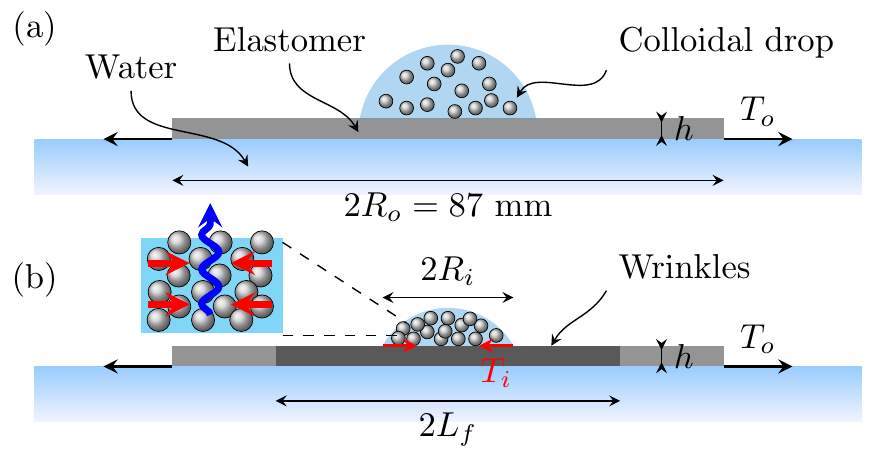}
    \caption{Schematic of the experimental set-up. A thin elastomeric sheet floats on a water bath.
        (a) The initial state where no wrinkles are observed after the drop deposition.
        (b) The final consolidated state where wrinkles extend over a total length $2 L_f$.
        The picture illustrates the drying stress induced by the Darcy flow in the porous material.
    }\label{fig:setup}
\end{figure}

In this Letter, we focus on visualizing and measuring the stress induced by the drying of a colloidal drop on a floating membrane (Fig.~\ref{fig:setup}).
The mechanical properties of the membrane are chosen carefully to satisfy two conditions.
First, the surface tension of the pure liquid drop alone and its weight do not  trigger the wrinkling instability (Fig. \ref{fig:setup}(a)).
Second, the drying-induced consolidation of the colloidal droplet, which induces a larger tensile stress in the membrane, triggers the instability (Fig. \ref{fig:setup}(b)).
From the length of the wrinkles, we can deduce the tension in the film.
Our aim is to examine the effect of particle size and polydispersity on the net stress that a drying drop of silica nanoparticles applies on the substrate.

Elastomeric sheets are prepared with polydimethylsiloxane (PDMS) by a spin coating technique on a silicon wafer (see Section S1 in Supplementary Information (SI) \cite{SI}).
The measured film thickness ranges from $[15.8,63.9]$ $\mu$m, the elastic modulus of the floating film is $E_s=1.2$ MPa and the Poisson ratio $\nu=0.5$.
Thus, the bending modulus ${\cal B} = \frac{E_s h^3 }{ 12 (1-\nu^2) }$ of the thin sheet is $ {\cal B}\in [5.3,350] \times 10^{-10}$ Pa $\cdot$ m$^3$.
The floating film is detached from the wafer in a water bath, which has a surface tension $\gamma_b = 55\pm 5$ mN/m.

The colloidal suspensions are silica nanoparticles.
SM, HS and TM are Ludox suspensions (purchased from Sigma-Aldrich), and Kleb and Lev denote Klebosol 50R50 and Levasil30, respectively (Table S1).
To ensure that all suspensions have the same pH, ionic strength and volume fraction, each suspension is dialysed \cite{Bouchoux2009} and has an initial volume fraction $\phi_0 = 0.15$.
The particle size distributions are characterized by Dynamic Light Scattering (DLS) and the average particle diameters $2a_0$ span from $7$ to $92$ nm.
Details on the sample preparation and on the particle size characterisation are presented in Section S2 in the SI \cite{SI}.
A drop of colloidal suspension of $8$ $\mu\ell$ is dispensed with a micropipette in the center of the floating sheet (Fig. \ref{fig:setup}).
Pictures of the floating film are recorded in a humidity controlled chamber ($R_H = 50$ \%) every minute.

\begin{figure}
    \centering
    \includegraphics[width=.8\linewidth]{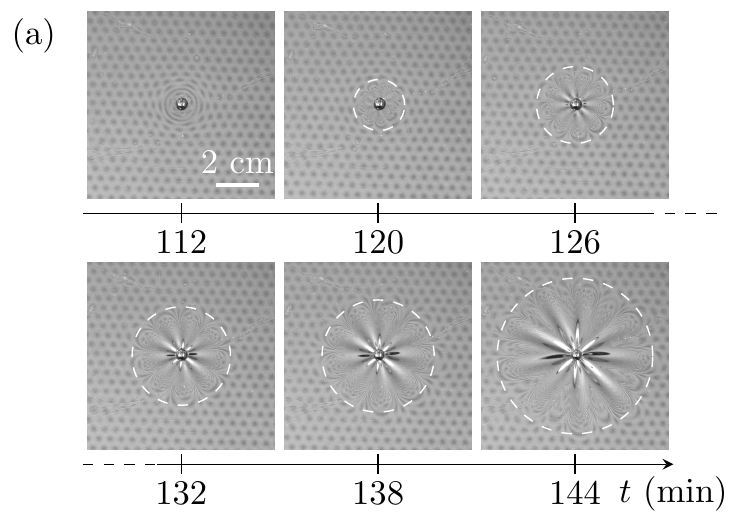}\\
    \includegraphics[width=1\linewidth]{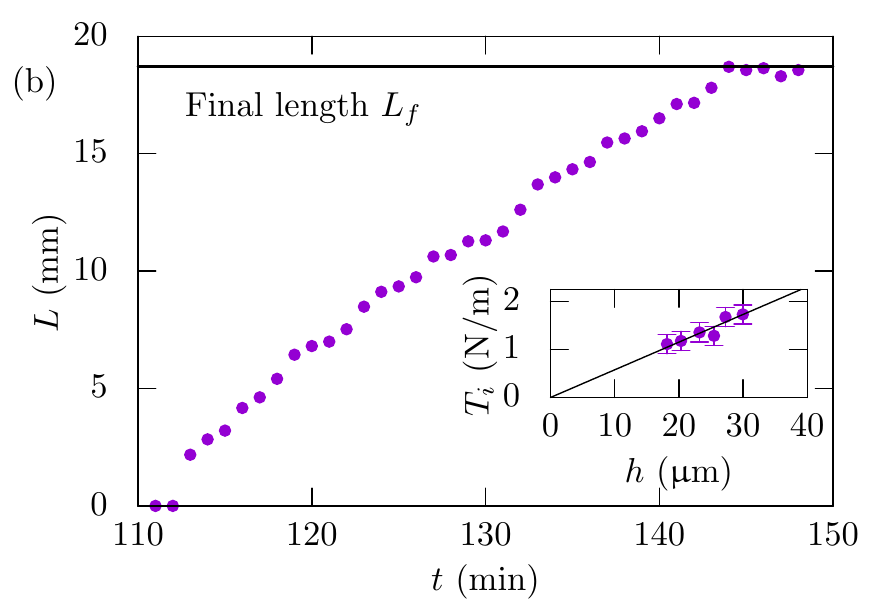}
    \caption{(a) Illustration of the expansion of wrinkles for a $8$ $\mu\ell$ drop of SM particles on a film of thickness $h=20$ $\mu$m.
        (b) The main plot shows the time evolution of the length $L(t)$ measured on the pictures shown above.
        The inset represents the inner tension $T_i$ deduced from Eq. (\ref{eq:FFT}) for different film thicknesses $h$ and for TM particles.
    }\label{fig:tension}
\end{figure}

Initially, no wrinkles are present on the floating film.
As the colloidal drop dries, the contact line first exhibits a stick-slip motion.
Ultimately, the contact line pins, the deposit height decreases and, concurrently, the wrinkling instability appears at $t_i=110\pm 15$ min (Fig. \ref{fig:tension}(a)).
The length of the wrinkled region grows linearly in time until it reaches a final length $L_f$ when the colloidal drop becomes solid at $t_f=150\pm 15$ min (Fig.~\ref{fig:tension}(b)).
During the consolidation, no relaxation of the wrinkles is observed, which indicates that the deposit is in good adhesion with the substrate (Fig. \ref{fig:tension}).
This observation is supported by a side view showing that the contact line is pinned during the development of the wrinkling instability and the drop height decreases (see Movie in SI \cite{SI}).
Furthermore, no cracks are visible in the final colloidal deposit, which is due to the substrate elasticity \cite{Smith2011}.

To obtain an estimate of the time evolution of the volume fraction, the time derivative of the liquid volume is ${\rm d} V(t)/{\rm d}t = -j$ where $j$ is the evaporative flux \cite{Deegan2000a}.
To estimate this flux, we consider that the aggregation occurs at $\phi_f = \phi(t_f)  \approx 0.6$ \cite{Boulogne2014a}.
Thus, we have $j \approx V_0 (1-\phi_0) / t_f$ and $V(t) \approx (1-\phi_0) V_0 - j t$.
%Therefore, the evaporation speed is $v_e \approx j/ (\pi R_i^2) \approx 2\times 10^{-7}$ m/s.
Therefore, the evaporation speed scales as $v_e \sim j/ S \approx 1\times 10^{-7}$ m/s where $S\approx  2 \pi R_i^2$ is the characteristic evaporating area.
The instability appears for $\phi(t_i) \approx 0.37$, which is the typical volume fraction for which the material becomes visco-plastic \cite{DiGiuseppe2012,Boulogne2014a}.

As we observed, the growth of the wrinkle length is associated with the consolidation of the deposit, which applies a tensile stress on the membrane.
We employ the final length of wrinkles to quantify this stress.
Davidovitch \textit{et al.} predicted the wrinkle's length as a function of the inner tension $T_i$ applied by the deposit at $r=R_i$ and the outer tension $T_o = \gamma_b$ due to the surface tension of the liquid bath at $r=R_o$ (Fig. \ref{fig:setup}) \cite{Davidovitch2011}.
The wrinkles morphology is set by the tension ratio $\tau = T_i/T_o$ and the bendability $\epsilon^{-1} = R_i^2 T_o / {\cal B}$.
With a stability analysis of the F\"oppl-von K\`arm\`an equations for $\epsilon\ll 1$, the length of the wrinkles is predicted to be
\begin{equation}\label{eq:NT}
    L_{NT} =  R_i \sqrt{T_i/T_o - 1}.
\end{equation}
This relation is valid for infinitesimal deformation amplitudes $\zeta$ and is called the Near Threshold limit (NT).

A second limit was also established for finite deformation amplitudes $\zeta$.
For $\epsilon \rightarrow 0$, Davidovitch \textit{et al.} assume that in the post-buckling region, the hoop stress $\sigma_{\theta\theta}$ and the shear stress $\sigma_{r\theta}$ vanish.
In this Far From Threshold limit (FFT), the wrinkle length is predicted to be
\begin{equation}\label{eq:FFT}
    L_{FFT} = \frac{R_i T_i}{2 T_o}.
\end{equation}

In our experimental conditions, the total number of wrinkles is established early in the development of the wrinkles as previously observed in wetting experiments \cite{Toga2013}.
Also, this number varies weakly, typically between 6 and 12, as the physical parameters are changed, which does not allow a prediction from this quantity described by scaling laws in the literature \cite{Davidovitch2011,King2012,Toga2013}.
From the final length of wrinkles $L_f$, the inner tension can be estimated in NT and FFT limits from equations (\ref{eq:NT}) and (\ref{eq:FFT}), respectively.
In our experimental conditions, the $T_i^{NT}$ values are about an order of magnitude above the range of NT tension values predicted by Davidovitch \textit{et al.} \cite{Davidovitch2011} (Section S4, SI \cite{SI}).
Consequently, inner tension values are calculated in the FFT limit from Eq. (\ref{eq:FFT}).

We performed experiments with a TM suspension, a drop volume of 8 $\mu\ell$, and different film thicknesses.
The tension $T_i$ resulting from the final wrinkle length (Eq. (\ref{eq:FFT})) is plotted as a function of the film thickness $h$ in the inset of Fig. \ref{fig:tension}(b).
The linear relationship between $T_i$ and $h$ suggests that the relevant parameter to describe the consolidation of the deposit on the membrane is the average stress $\sigma_i = T_i / h$.

Since the deposit has a good adhesion on the elastomer \cite{Boulogne2014b}, the drying strain in the deposit is transmitted to the flexible membrane, which leads to the formation of wrinkles.
In the following, we consider that the deposit generates an effective stress $\sigma_i$ in the membrane located at the deposit edge.

\begin{figure}
    \centering
    \includegraphics[width=.90\linewidth]{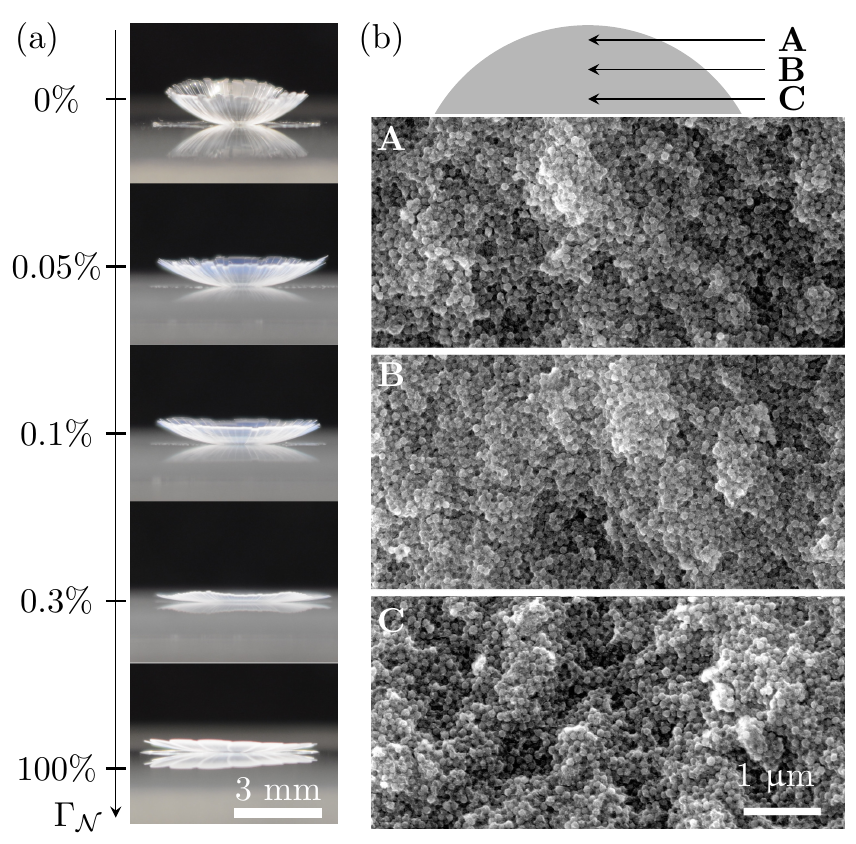}
    \caption{
        (a) Delamination of colloidal drops upon drying on a glass slide for different number ratios $\Gamma_{\cal N}$.
        Pictures are taken at the maximum elevation.
        The initial volume is 6 $\mu\ell$ and the initial volume fraction of the suspensions is 0.15.
        (b) Scanning-electron-microscope (SEM) images in the cross-section of a dried deposit at a number ratio $\Gamma_{\cal N} = 0.003$.
        Positions are indicated by the schematic.
    }\label{fig:final}
\end{figure}

To study the effect of the particle size polydispersity \cite{Iler1979,Lidon2014} on the drying stress, we prepared blended suspensions from SM and Lev samples (Table S1).
Since Lev particles are larger than the largest particles present in the SM suspension, we consider as a first approximation that the mixture is a bidisperse suspension; the size ratio is approximately a factor of 10.
We denote $a_s$ and $a_\ell$, the average radii of small and large particles, respectively.
The volume ratio of large particles is defined as
\begin{equation}\label{eq:gammav}
    \Gamma_{\cal V} = \frac{{\cal V}_{\ell}}{{\cal V}_{s} + {\cal V}_{\ell}},
\end{equation}
where ${\cal V}$ is the volume of particles.
Equivalently, denoting ${\cal N}_s$ and ${\cal N}_\ell$ as the number of small and large particles, we define the particle number ratio of large particles as
\begin{equation}\label{eq:gamman}
    \Gamma_{\cal N} = \frac{{\cal N}_{\ell}}{{\cal N}_{s} + {\cal N}_{\ell}} = \left[ 1 + \frac{a_\ell^3}{a_s^3} \left(\Gamma_{\cal V}^{-1} - 1\right) \right]^{-1}.
\end{equation}

We conducted a first set of experiments that consist in drying drops of colloids on a hydrophilic glass slide.
The drying occurs from the contact line toward the center as the edge is thinner and the evaporative flux is highest at the edge.
During the consolidation, the material delaminates from the substrate and the curvature increases with the tensile stress \cite{Pauchard2006}.
In Fig. \ref{fig:final}(a), we observe that the addition of a small number of large particles reduces significantly the delamination, which could be crucial for coating applications.
Next, we investigate the stress reduction caused by the presence of these large particles.

Since measurements of the drying stress are possible with the floating film (Fig. \ref{fig:tension}), we quantify the effect of the particle size ratio.
As the evaporation may lead to particle segregation in the deposit \cite{Trueman2012,Routh2013,Fortini2016}, we checked the particle organization across the thickness of the deposit by Scanning Electron Microscopy.
We observed that the particles remains homogeneously distributed at the end of the drying (Fig. \ref{fig:final}(b) and SI, Section S3 \cite{SI}).

\begin{figure}
    \centering
    \includegraphics[width=.9\linewidth]{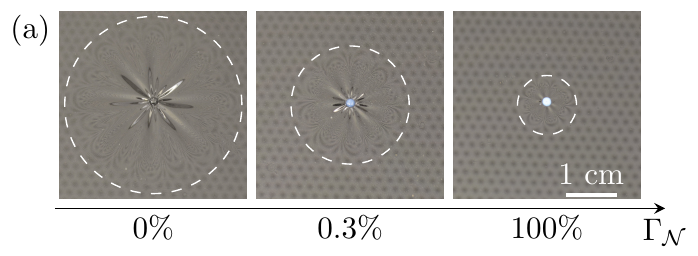}\\
    \includegraphics[width=1\linewidth]{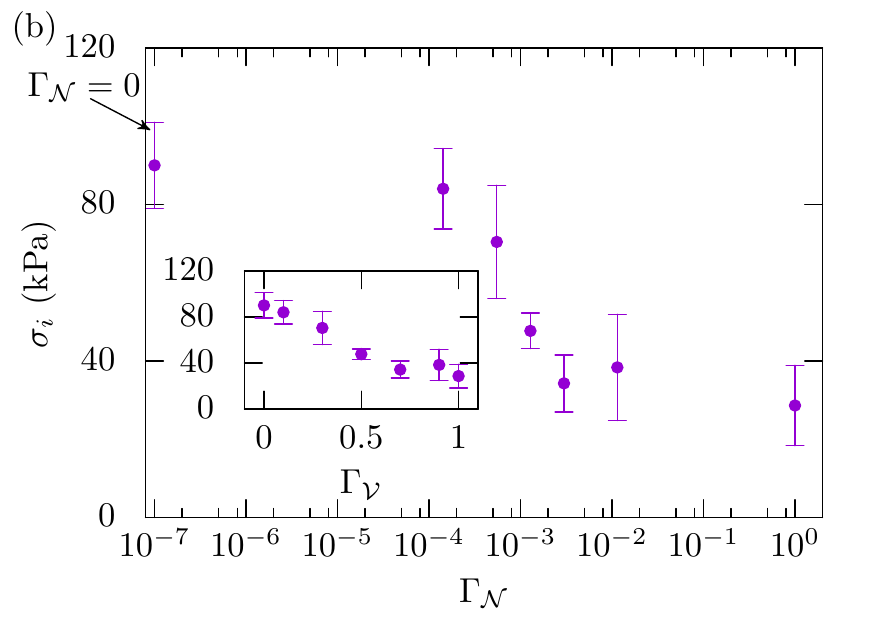}
    \caption{
        (a) Pictures of the final state for three different number ratios $\Gamma_{\cal N}$ (Eq. \ref{eq:gamman}).
        (b) Final tensile stress $\sigma_i$ applied by a deposit on a membrane for a bidisperse suspension indicated by the number ratio of large particles $\Gamma_{\cal N}$.
        The inset shows the same data as a function of $\Gamma_{\cal V}$ (Eq. \ref{eq:gammav}).
    }\label{fig:blend}
\end{figure}

Our experiments showed that the length of the wrinkles decreases as the number fraction of large particles increases (Fig. \ref{fig:blend}(a)).
The stress $\sigma_i = T_i^{FFT}/h=2 T_o L_{FFT} / (R_i h)$ after drying is reported in Fig. \ref{fig:blend}(b) as a function of $\Gamma_{\cal N}$ and $\Gamma_{\cal V}$.
First, we find that the drying stress for a pure suspension of large particles ($\Gamma_{\cal N}=1$) is about half as large as the case of a pure suspension of small particles ($\Gamma_{\cal N}=0$), which confirms that the final stress depends on the particle size.
These results also show that the addition of a number density $\Gamma_{\cal N}\approx 0.003$ of large particles ($\Gamma_{\cal V}\approx 0.7$) leads to a stress value comparable to a suspension of only large particles.

To interpret these results, we refer to the description developed by Sherwood \cite{Sherwood1929} and further detailed in particular by Brinker and Scherer \cite{Brinker1990}.
During the drying of the colloidal suspension, the particles concentrate and form a porous material for which the surface remains wet.
The liquid flow driven by the evaporation generates a pressure gradient due to the Darcy law, which is responsible for a tensile stress in the deposit \cite{Brinker1990,Boulogne2015}.
When the compression exceeds the repulsive interactions between the colloids, silica particles irreversibly aggregate \cite{Iler1979} and the menisci penetrate in the material \cite{Thiery2015}.

\begin{figure}
    \centering
    \includegraphics[width=1\linewidth]{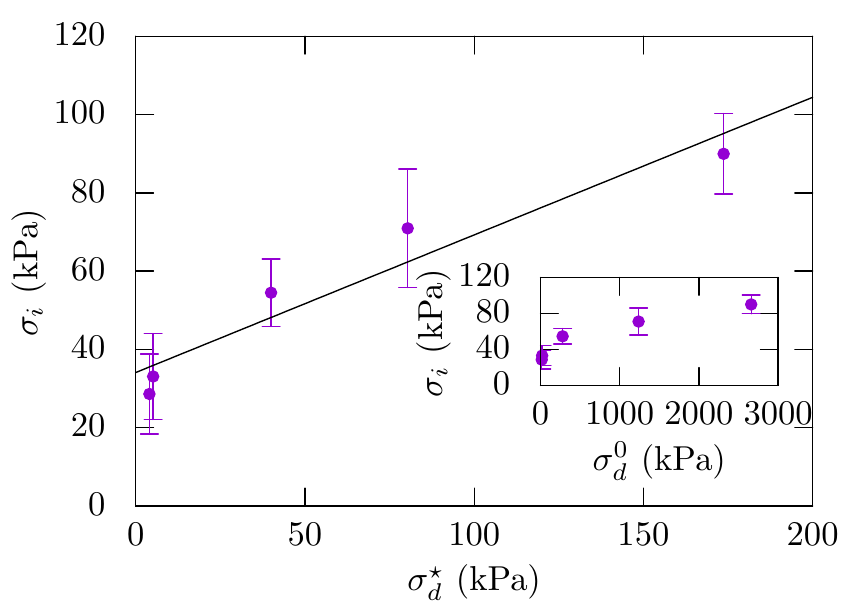}
    \caption{
        Stress $\sigma_i = T_i/h$ as a function of the Darcy stress $\sigma_d$ (Eq. \ref{eq:darcy}).
        In the main plot, $\sigma_d^\star$ is calculated for the permeability $k(a^\star)$ where the particle radius $a^\star$ is representative of the large particles size of the distribution.
        The inset shows the same results for $\sigma_d^0$ related the average particle size $a_0$.
        The solid line is a guide for the eye.
    }\label{fig:stress}
\end{figure}

We use the Darcy approximation $\eta {\bm v}/k(a) = -{\bm \nabla} p$, where $\eta=10^{-3}$ Pa$\cdot$s is the solvent viscosity, ${\bm v}$ is the fluid velocity, $p$ is the pressure and $k(a)$ is the permeability.
We can estimate the tensile stress $\sigma_d$ from the pressure gradient in the porous network as
\begin{equation}\label{eq:darcy}
    \sigma_d \sim \frac{ R_i \eta v_e}{k(a)},
\end{equation}
where $R_i$ is the deposit size, and $v_e$ is the evaporation speed.
For a characteristic particle size $a=25$ nm and $R_i=0.8$ mm, we have $\sigma_d\approx 30$ kPa, which is comparable to the measured $\sigma_i$.
For a monodisperse suspension of particle radius $a$, the permeability depends on the square of the pore size and can be calculated by the Carman-Kozeny relation
$k(a)=\frac{a^{2} (1-\phi_f)^3}{45 \phi_f^2}$, where $\phi_f$ is the final volume fraction.
We observed that the large particles dominate the stress even if they are present at a low number fraction (Fig. \ref{fig:blend}(b)).
We interpret this observation as indicating the formation of a continuous path for flow due to large pore sizes formed by the large particles, which increases the permeability.

We performed experiments for different average particle sizes.
The results are represented in Fig. \ref{fig:stress} and the inset shows the stress $\sigma_i$ calculated from the wrinkle length as a function of the Darcy stress $\sigma_d^0$ for the average particle size $a_0$ (Eq. \ref{eq:darcy}).
For the permeability, the Carman-Kozeny relation is used to calculate a characteristic value.
Moreover, we define a particle size $a^\star$ to take into account the particle polydispersity.
This size is defined as the average particle size of the $30$\% largest particles of the distribution (Section S2, SI \cite{SI}).
From these values, we calculate the Darcy stress $\sigma_d^\star$ shown in Fig. \ref{fig:stress}.
These results indicate that the stress measured from the length of the wrinkles is comparable to $\sigma_d^\star$, while it is significantly different if the average particle size is considered.
Thus, we conclude that the large particles plays a major role in the final drying stress of a colloidal suspension.
The detailed relationship between the drying stress and the stress in the membrane remains to be established and has to account for the geometrical aspects of the deposit and the deformation of the membrane beneath the deposit.

In summary, we have studied the consolidation of colloidal silica nanoparticles in the  geometry of a drop placed in the center of a thin circular elastomeric sheet floating on a liquid bath.
As the drying proceeds, the consolidation of the colloidal drop triggers a wrinkling instability of the membrane.
We demonstrated that this technique allows a measurement of the final tensile stress applied by the deposit on the substrate.
We used this method to measure the effects of the particle size and the polydispersity.

For a mixture of small and large particles with a typical size ratio of about 10, the addition of about 1 \% of large particles reduces the stress value obtained for the suspension of small particles to the value observed for large particles only.
Indeed, colloidal suspensions are generally polydisperse.
Our work provides evidence that the contribution of the particle size polydispersity on the tensile stress is dominated by the largest particles.
We interpreted these results by the percolation of the largest particles that creates paths with a larger pore size, which reduces the tensile stress due to the pressure gradient of the Darcy flow, accompanying evaporation of the liquid.

We believe that these results are significant for future investigations regarding the quantitative measurement of the drying stress in more complex materials.
Our findings could influence the study of the emergence of tensile stress in drying materials, which often leads to delamination or substrate deformations that must be avoided in many technological applications.
As future work, it will be important to derive a numerical or theoretical model to understand more quantitatively the effect of the polydispersity on the tensile stress.

%and the Imaging \& Analysis Center at Princeton University
%\paragraph{Acknowledgements}
\begin{acknowledgments}
We are very grateful to I. Cantat, B. Davidovitch, B. Saintyves and J.-B. Salmon for stimulating discussions.
We thank J.-M. Copin and M. Persson for providing Klebosol and Levasil samples, respectively.
We also thank R. K. Prud'homme for the use of the Malvern Zetasizer, J. Feng and B. K. Wilson for assistance, as well as J. Schreiber  for assistance with SEM images.
F.B. acknowledges that the research leading to these results received funding from the People Programme (Marie Curie Actions) of the European Union's Seventh Framework Programme (FP7/2007-2013) under REA grant agreement 623541.
\end{acknowledgments}

%\bibliography{article_flower}
%\bibliographystyle{unsrt}

%%%%%%%%%%%%%%%%%%%%%%%%%%%%%%%%

\newpage
\clearpage
\begin{center}
\Large Supporting Information
\end{center}

\beginsupplement

\setcounter{page}{1}

\section*{S1. Detailed experimental setup}

\paragraph{Floating films} Polydimethylsiloxane (PDMS, sylgard 184, Dow Corning) is mixed and degassed at $0^\circ$C to slow down the reaction and to ensure a reproducible viscosity.
A silicon wafer is first coated with a 10\% wt. aqueous solution of polyvinyl alcohol (PVA, Sigma-Aldrich, $M_w = 10$ kDa).
The coating is dried until a uniform color is obtained.
Then, the PDMS is spin coated on this wafer and placed in an oven at $65^\circ$C for 10 hours.
The edge of the coated PDMS is removed to obtain a circular film of diameter $87$ mm.
A petri dish, painted in black to enhance the contrast on the picture, is partially filled with deionized water.
The wafer is dipped in this bath to dissolve the sacrificial layer of PVA and to detach the PDMS film.
Once the PDMS film is fully detached and floats at the surface, the wafer is removed.
We used a white light spectrometer (OceanOptics USB2000+ used with a LS-1-LL tungsten halogen light source) to measure the film thicknesses, which range between $[15.8,63.9]$ $\mu$m.
The presence of PVA lowers the surface tension of the bath, which is $\gamma_b = 55\pm 5$ mN/m.

\begin{figure*}[h!]
    \centering
    \includegraphics[width=0.99\linewidth]{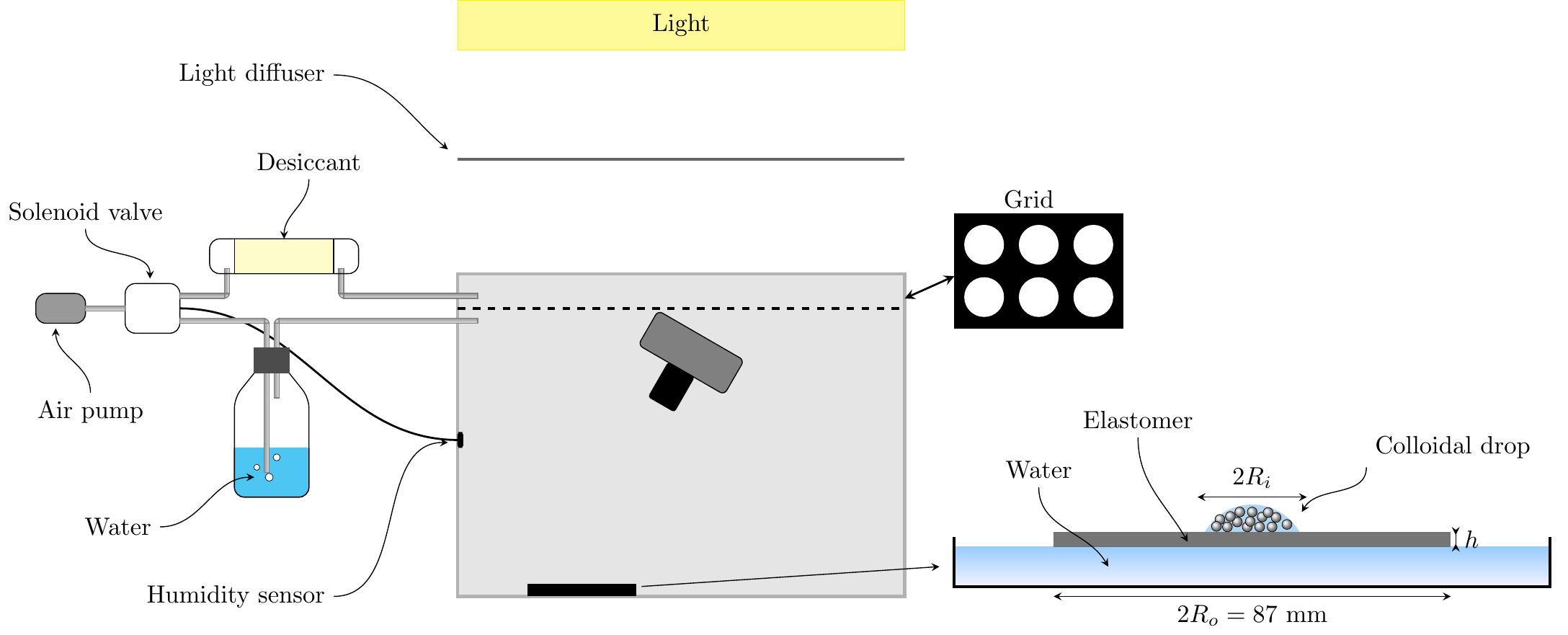}
    \caption{Detailed experimental setup.
        A grid is placed between the light and the sample.
    The reflection of the grid on the bath interface allows the visualisation of the deformation of the elastomer.}\label{fig:SI_setup}
\end{figure*}

\paragraph{Visualisation}
Our experiments of controlled drying of drops are performed in a transparent glove box, which has humidity controller presented in Fig. \ref{fig:SI_setup}.
The relative humidity is set to $50$\% in all of our experiments.
A grid is placed in the glove box on the top of the sample and a light source is placed behind a light diffuser to obtain an even illumination.
The deformation of the membrane is visualised by the distortion of the pattern made by the grid.
The time evolution of the system is recorded by a Nikon camera (D7100) mounted with a 18-55mm Nikon objective (Fig.~\ref{fig:SI_setup}).

\section*{S2. Preparation and characterization of the colloidal suspensions}

\paragraph{Dialysis}

Each suspension was dialysed to ensure that they have the same properties (ionic strength and pH).
Dialysis bags (12-14 kDa, Fisherbrand) are partially filled with the colloidal solutions.
A solution denoted s1 is prepared from deionized water, the pH is raised to 9.5 by the addition of NaOH and the ionic strength is adjusted by the addition of $5$ mM of NaCl.
To wash these suspensions, the bags are plunged in large baths of solution s1 for 10 days and baths are replaced twice.
Then, the diluted suspensions are concentrated by osmotic stress.
This second step consists in preparing a bath in which  polyethylene glycol (Sigma-Aldrich, 35 kDa) is dissolved in a solution s1.
Mass fractions are measured and the suspensions are diluted with a solution s1 to reach a mass fraction $\Phi_0=0.25$, which corresponds to a volume fraction $\phi_0=0.15$.

\paragraph{Particle size}
The particle size distributions are characterized by Dynamic Light Scattering (DLS) with a Malvern Zetasizer Nano ZS instrument.
Samples are diluted to a volume fraction of $\approx 0.001$.
We denote $n(a)$ the distribution function of particle number and $v(a)$, the distribution function in volume, and satisfy
\begin{eqnarray}
    \int_0^\infty n(a'){\rm d}a'=1\\
    \int_0^\infty v(a'){\rm d}a'=1
\end{eqnarray}
The average particle radius $a_0$ is defined as
\begin{equation}
    a_0 = \int_0^\infty a' n(a') {\rm d} a'.
\end{equation}
The particle radius $a^\star$, representative of the large particles of the distribution is defined as follow.
We consider 30\% of the tail of the particle distribution $v(a)$, \textit{i.e.} $a>a_c$ with
\begin{equation}
    \int_0^{a_c} v(a') {\rm d} a' = 0.7.
\end{equation}
The radius $a^\star$ is the average particle size of the distribution tail, \textit{i.e}
\begin{equation}
    a^\star = \int_{a_c}^\infty a' n(a') {\rm d} a'.
\end{equation}
The suspension properties are summarized in Table \ref{tab:particles}.

\begin{minipage}{.5\textwidth}\centering
    \includegraphics[width=0.95\linewidth]{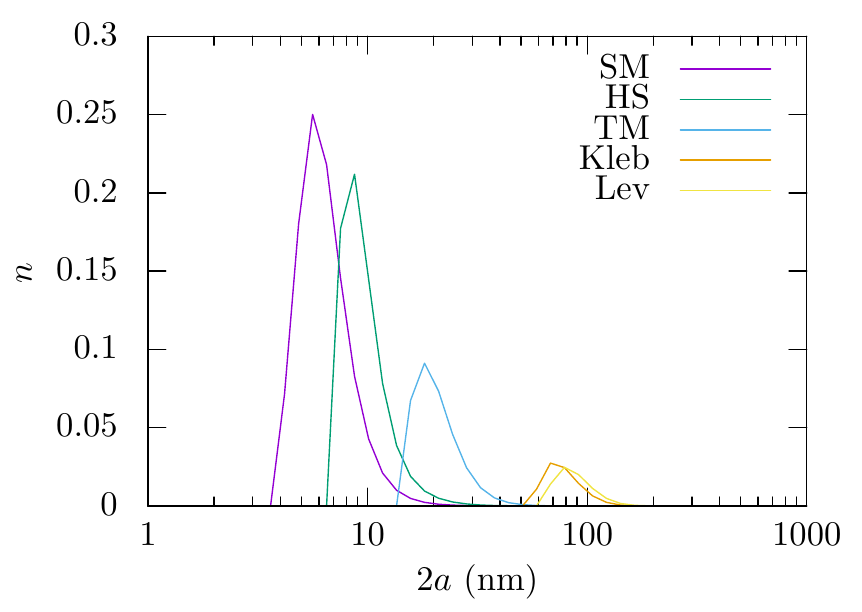}
    \captionof{figure}{Particle distribution function $n(a)$ for the number of particles.}\label{fig:DLS}
\end{minipage}
\begin{minipage}{.5\textwidth}\centering
    \begin{tabular}{|c|c|c|c|c|c|c|}
        \hline
        Name & SM & HS & TM & Kleb & Lev  \\
        \hline
        $2\,a_0$ (nm) & 7 & 11 & 23 & 81 & 92 \\
        $2a^\star$ (nm) & 20 & 30 & 42 & 117 & 132 \\
        \hline
    \end{tabular}
    \captionof{table}{Properties of the colloidal suspensions of silica nanoparticles. For all suspensions, the pH is 9.5, the ionic strength is $[$NaCl$]$= 5 mM and the volume fraction is $\phi_0=0.15$.
    }
    \label{tab:particles}
\end{minipage}

\section*{S3. Particle organization}

For blend suspensions, we define the particle volume ratio as
\begin{equation}\label{eq:gammav}
    \Gamma_{\cal V} = \frac{{\cal V}_{\ell}}{{\cal V}_{s} + {\cal V}_{\ell}},
\end{equation}
and the particle number ratio as
\begin{equation}\label{eq:gamman}
    \Gamma_{\cal N} = \frac{{\cal N}_{\ell}}{{\cal N}_{s} + {\cal N}_{\ell}} = \left[ 1 + \frac{a_\ell^3}{a_s^3} \left(\Gamma_{\cal V}^{-1} - 1\right) \right]^{-1},
\end{equation}
where $a_s$ and $a_\ell$ are the average radii of small and large particles, respectively.

SEM images of dried drops of colloidal suspensions were captured with the FEI Quanta 200 FEG Environmental-SEM in low vacuum mode.
Samples are prepared in the same conditions as presented in section S1.
To image cross-sections, a razor blade is gently pressed on the top of the deposit to nucleate a crack along the diameter.
The two pieces are then separated and placed on a sample holder.

\begin{figure*}[h!]
    \centering
    \includegraphics[width=.8\linewidth]{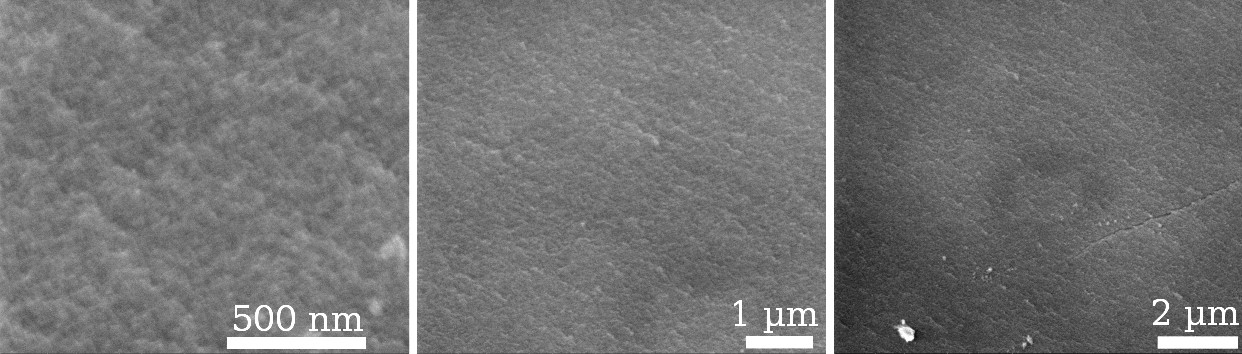}
    \caption{
        Scanning-electron-microscope (SEM) images in the cross-section of a dried deposit of SM particles ($\Gamma_{\cal N}=0$).
        The different pictures show different magnifications.
    }\label{fig:SI_sem_sm}
\end{figure*}

\begin{figure*}[h!]
    \centering
    \subfloat[Base]{\includegraphics[width=4.0cm]{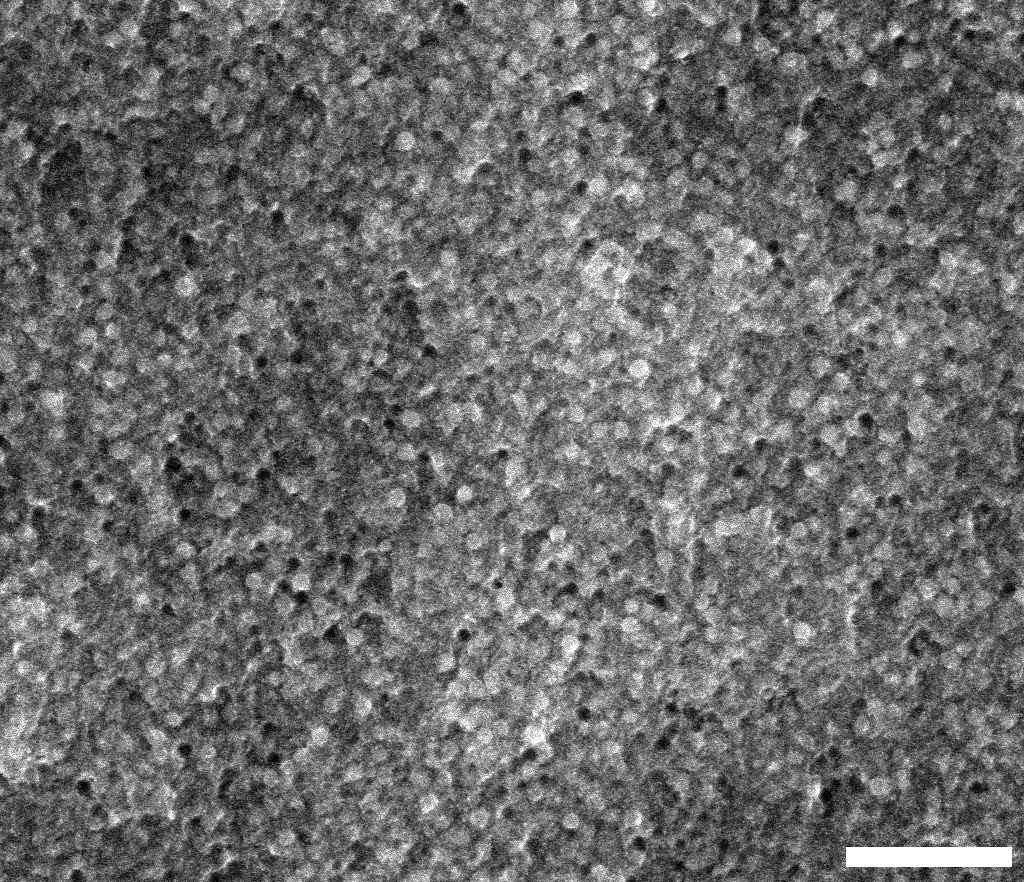}\label{subfig:thick}} \hspace{1cm}
    \subfloat[Middle]{\includegraphics[width=4.0cm]{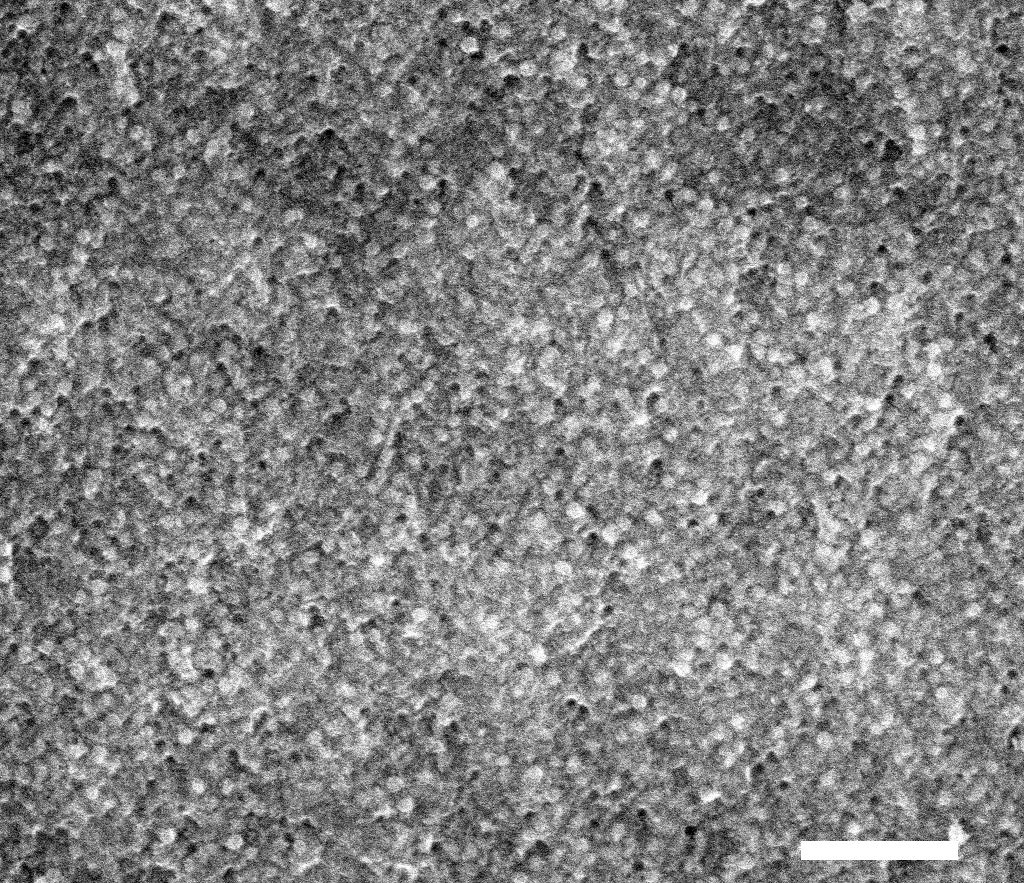}\label{subfig:thick}} \hspace{1cm}
    \subfloat[Top]{\includegraphics[width=4.0cm]{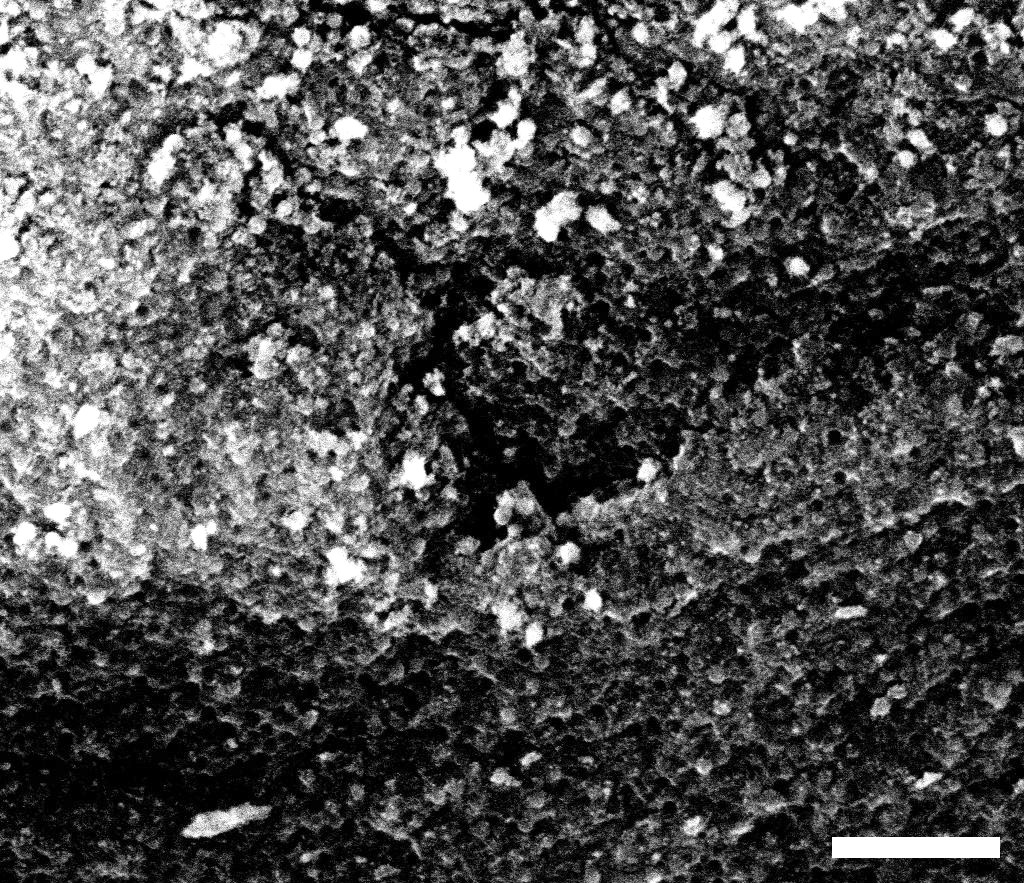}\label{subfig:thick}} \hspace{1cm}
    \subfloat[Near the contact line]{\includegraphics[width=4.0cm]{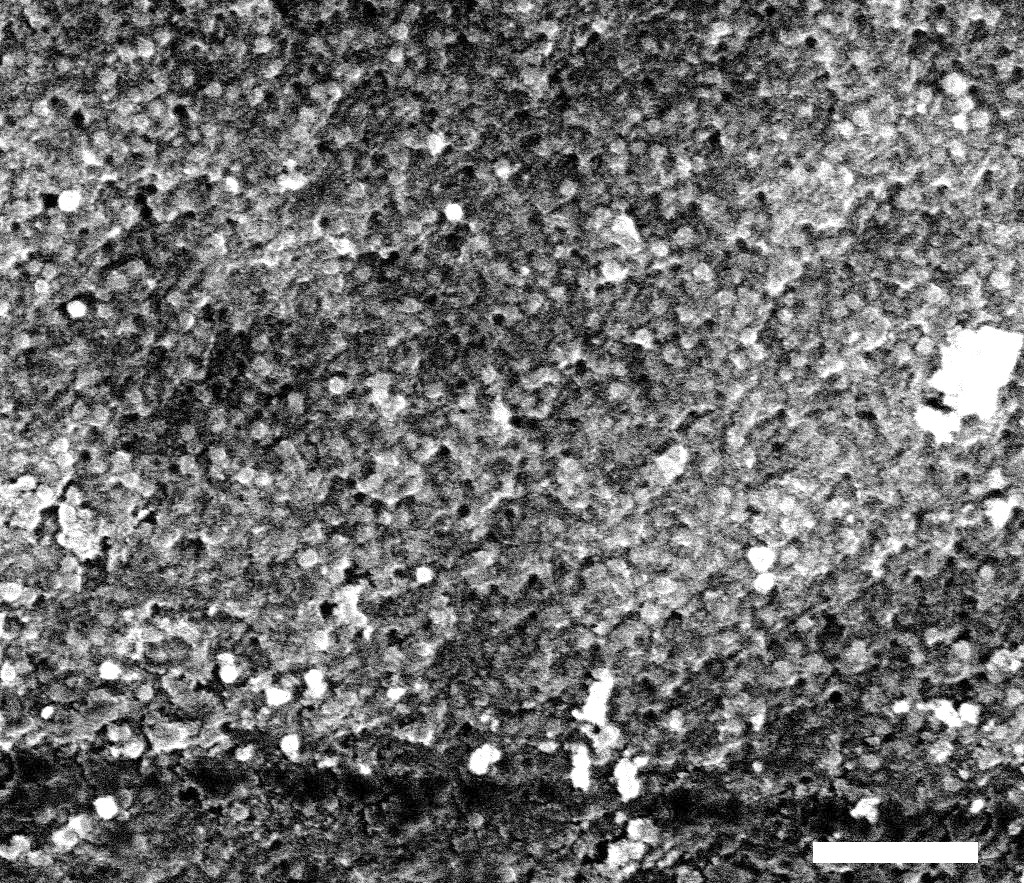}\label{subfig:thick}} \hspace{1cm}
    \caption{
        Scanning-electron-microscope (SEM) images in the cross-section of a dried deposit for a number ratio $\Gamma_{\cal N}=0.05$\%.
        Images are taken at different locations and the scale bars represent 1 $\mu$m.
    }\label{fig:SI_sem_30pc}
\end{figure*}

\begin{figure*}[h!]
    \centering
    \includegraphics[width=1\linewidth]{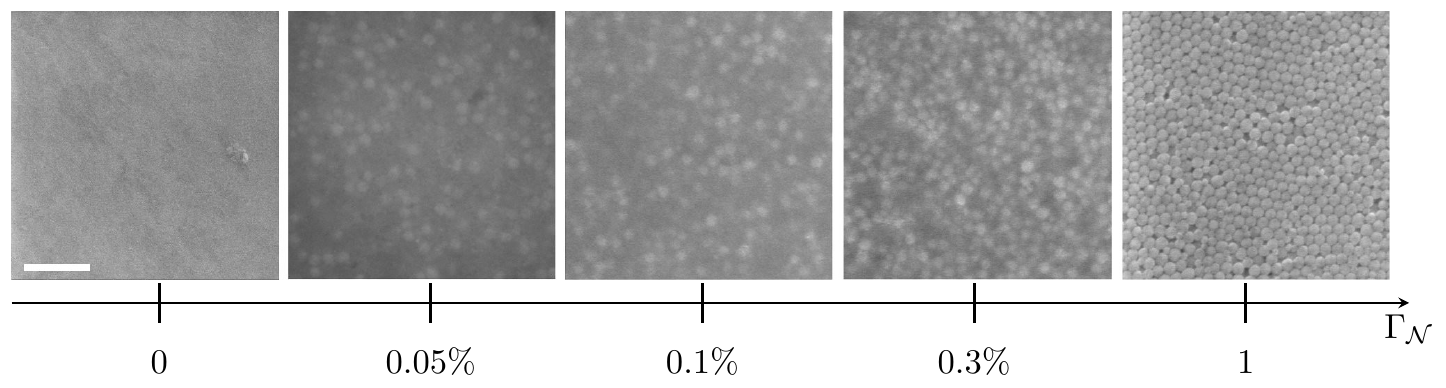}
    \caption{
        Scanning-electron-microscope (SEM) images of the colloidal deposit surface after complete drying for different  number ratios of particles $\Gamma_{\cal N}$.
        Only the large particles can be imaged.
        The scale bar represents $500$ nm.
    }\label{fig:SI_sem_surface}
\end{figure*}

In this section, we present additional images.
In Fig. \ref{fig:SI_sem_sm}, the cross section of a dried deposit containing only SM particle is imaged and we see that the particle size is smaller than the resolution we could obtain.
In Fig. \ref{fig:SI_sem_30pc}, we present cross sections images for $\Gamma_{\cal N}=0.05$\%.
As for the pictures in the main text, we do not see a segregation of the large particles across the thickness.
Also, similar particle organisations are obtained on the top and near the contact line.
In Fig. \ref{fig:SI_sem_surface}, we imaged the surface of the deposits for different $\Gamma_{\cal N}$ where we notice the presence of the large particles as their concentration increases.

\textcolor{black}{
In our experiment, we can estimate the P\'eclet numbers for small (${\rm Pe}_s$) and large (${\rm Pe}_\ell$) particles.
The P\'eclet number is defined as ${\rm} Pe = \frac{v_e h}{D}$ where $v_e$ is the evaporation speed, $h$ the characteristic thickness and $D$ the diffusion coefficient.
We use the Stokes-Einstein diffusion coefficient $D=k_B T/(6\pi\eta a)$, where $k_B$ is the Boltzmann constant, $T$ is the absolute temperature, $\eta=10^{-3}$ Pa$\cdot$s is the solvent viscosity and $a$ the particle radius.
}

\textcolor{black}{
For our bidisperse suspensions and considering the average particle size of each species, we have ${\rm Pe}_s\approx1.6$ and ${\rm Pe}_\ell\approx 21$ for $h \approx 1$ mm.
Fortini \textit{et al.} evidenced numerically and experimentally a segregation effect in drying films with ${\rm Pe} \gg1$ for both species (Fortini {\textit et al.} PRL 116, 118301 (2016)).
However, as shown in our SEM images, we could not observe a significant segregation of the two species in the dried film.
Nevertheless, as pointed by Fortini \textit{et al.}, the segregation is a dynamic process that is observed after a transient state.
The duration of this transient must be function the specific interactions between the particles and the initial volume fraction.
In our system, the material becomes visco-plastic at a volume fraction about $0.3$ and our initial volume fraction is $0.15$.
Thus, a possible explanation is that the system cannot evolve to a segregated state.
This observations would deserve further investigations, which are beyond the scope of the present study.
}

\newpage
\clearpage
\section*{S4. Bendability-tension diagram}

\textcolor{black}{In our experiments, we measure the final length of the wrinkles $L_f$. Then, from models derived by Davidovitch \textit{et al.} \cite{Davidovitch2011}, we calculate the inner tension $T_i$ applied by the deposit on the film.}

\begin{figure*}%[h!]
    \centering
    \includegraphics[width=0.49\linewidth]{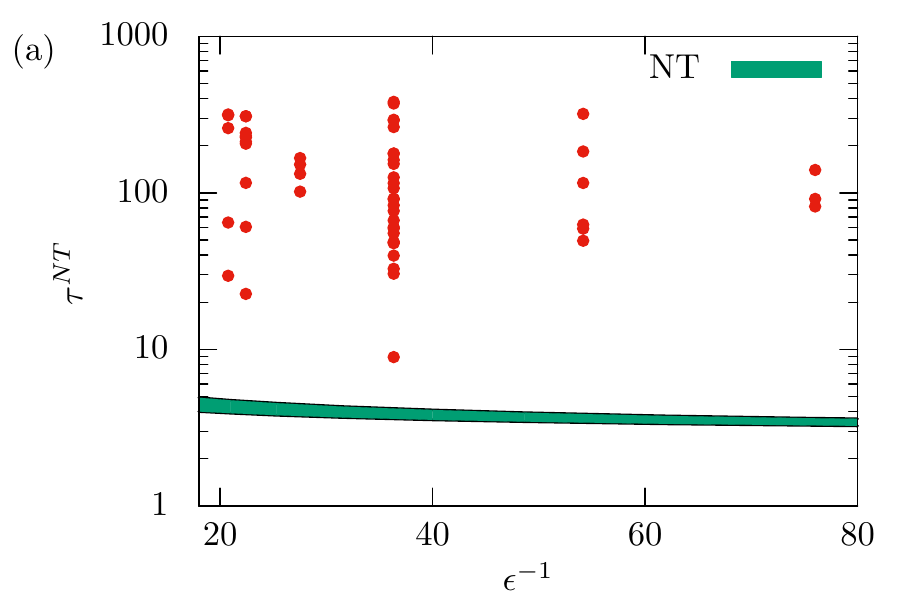}
    \includegraphics[width=0.49\linewidth]{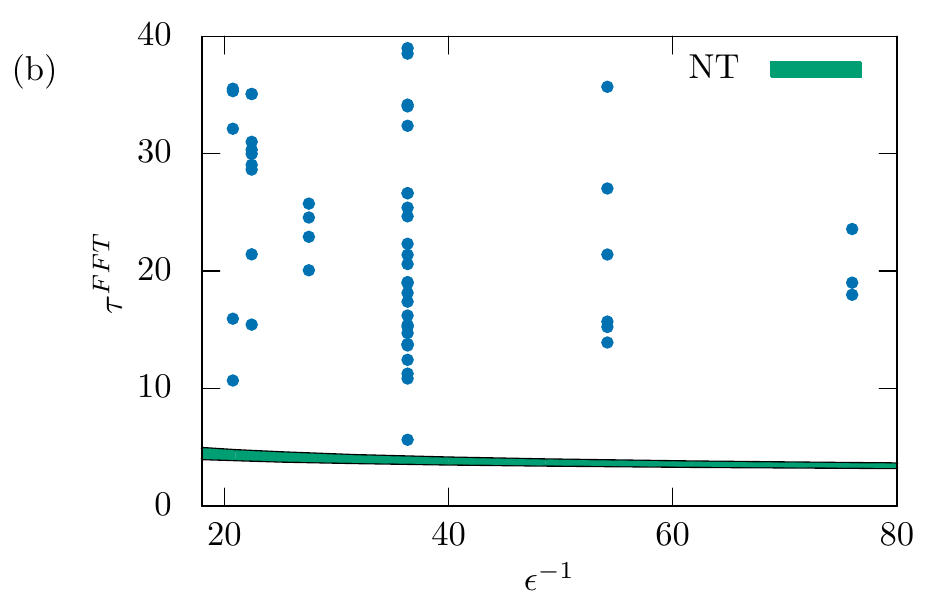}
    \caption{
        Tension ratio $\tau = T_i/T_o$ as a function of the bendability $\epsilon^{-1}$.
        The points correspond to our experiments  (a)  where we used the expression from the NT limit to calculate the tension ratio $\tau^{NT} = T_i^{NT}/T_o = 1 + L_f^2 / R_i^2$,
        and (b), we used the expression from the FFT limit $\tau^{FFT} = T_i^{FFT}/T_o = 2 L_f/R_i$.
        On both plots, the green band represents the Near Threshold (NT) domain predicted by Davidovitch \textit{et al.} \cite{Davidovitch2011}.
    }\label{fig:SI_diagram}
\end{figure*}

Davidovitch \textit{et al.} predict this tension in two different domains.
The first domain is named Near Threshold (NT) limit and correspond to the limit of infinitesimal deformation amplitude $\zeta$ and large bendability ($\epsilon^{-1} = R_i^2 T_o / {\cal B}\gg 1$.
In that domain, the tension ratio $\tau^{NT} = T_i^{NT}/T_o$ is given by
\begin{equation}\label{eq:NT}
	\tau^{NT} = 1 + \frac{L_f^2}{R_i^2},
\end{equation} 
which corresponds to equation (1) in the Letter. The tension ratios for this limit and for all our experiments are presented in Fig. \ref{fig:SI_diagram}(a).
The second domain is the Far From Threashold (FFT) limit and it assumes that the deformation $\zeta$ is finite and $\epsilon \rightarrow 0$. Therefore, the tension ratio is
\begin{equation}\label{eq:FFT}
	\tau^{FFT} = \frac{2 L_{f}}{R_i},
\end{equation}
which corresponds to equation (2) in the Letter.
Similarly, we plot this prediction for our data in Fig. \ref{fig:SI_diagram}(b).

In Fig. \ref{fig:SI_diagram}, we also represent the NT domain predicted by Davidovitch \textit{et al.} \cite{Davidovitch2011} and we observe that both estimations are above the NT limit.
If the NT limit would describe our results, we expect that $T_i^{NT}$ values to be in the NT green domain, but the prediction is about at least an order of magnitude higher.
Thus, the estimate from the NT limit is not self-consistent.
Let us now consider the values deduced from the FFT limit. 
These values are also above the NT domain (represented in green in Fig. \ref{fig:SI_diagram}(b)), which is self-consistent.
Also, we observe that $\tau^{NT} \approx \tau^{FFT}$, which further validates that the NT prediction is not valid as the prediction is significantly above the NT domain.
Consequently, for all our measurements, we considered $T_i = T_i^{FFT}$.

Let us remark that equations (\ref{eq:NT}) and (\ref{eq:FFT})
are valid for $R_i \ll R_o$. More general equations have been established recently \cite{Taylor2015} but they do not modify significantly the estimate of the tension in the film in our experiments, which confirms this assumption.

\end{document}